\documentclass[12pt,a4paper,final]{iopart}

\usepackage{iopams}  
\usepackage{graphicx}
\usepackage[breaklinks=true,colorlinks=true,linkcolor=blue,urlcolor=blue,citecolor=blue]{hyperref}

\expandafter\let\csname equation*\endcsname\relax
\expandafter\let\csname endequation*\endcsname\relax

\usepackage{amsmath}
\usepackage{amssymb}
\usepackage{amsthm}
\usepackage{float}
\usepackage[T1]{fontenc}
\usepackage[latin9]{inputenc}
\usepackage{geometry}
\usepackage{esint}
\usepackage{multirow}
\usepackage{xcolor}
\usepackage[normalem]{ulem}
\usepackage{lipsum}

\newcommand{\bra}[1]{\ensuremath{\left\langle#1\right|}}
\newcommand{\ket}[1]{\ensuremath{\left|#1\right\rangle}}

\graphicspath{{./Images/}}

\usepackage{mathtools} 

\usepackage{bm} 

\usepackage{verbatim} 

\newcommand{\be}{\begin{equation}}
\newcommand{\bea}{\begin{eqnarray}}
\newcommand{\ee}{\end{equation}}
\newcommand{\eea}{\end{eqnarray}}
\newcommand{\ben}{\begin{equation*}}
\newcommand{\bean}{\begin{eqnarray*}}
\newcommand{\een}{\end{equation*}}
\newcommand{\eean}{\end{eqnarray*}}
\newcommand{\ba}{\begin{align}}
\newcommand{\ea}{\end{align}}
\newcommand{\ban}{\begin{align*}}
\newcommand{\ean}{\end{align*}}

\begin{document}

\title[Ghost Imaging]{Ghost imaging with engineered quantum states by Hong-Ou-Mandel interference}

\author{Nicholas Bornman$^{1}$, Shashi Prabhakar$^{1,2}$, Adam Vall\'es$^{1,3,4}$, Jonathan Leach$^{5}$, Andrew Forbes$^{1}$}
\address{$^1$School of Physics, University of the Witwatersrand, Private Bag 3, Wits 2050, South Africa}
\address{$^2$CSIR National Laser Centre, P.O. Box 395, Pretoria 0001, South Africa}
\address{$^3$Graduate School of Advanced Integration Science, Chiba University, 1-33 Yayoi-cho, Inage-ku, Chiba 263-8522, Japan}
\address{$^4$Molecular Chirality Research Center, Chiba University, 1-33, Yayoi-cho, Inage-ku, Chiba 263-8522, Japan}
\address{$^5$IPaQS, SUPA, Heriot-Watt University, Edinburgh EH14 4AS, United Kingdom}

\ead{adam.vallesmari@wits.ac.za}

\begin{abstract}
Traditional ghost imaging experiments exploit position correlations between correlated states of light. These correlations occur directly in spontaneous parametric down-conversion (SPDC), and in such a scenario, the two-photon state usually used for ghost imaging is symmetric. Here we perform ghost imaging using an anti-symmetric state, engineering the two-photon state symmetry by means of Hong-Ou-Mandel interference. We use both symmetric and anti-symmetric states and show that the ghost imaging setup configuration results in object-image rotations depending on the state selected. Further, the object and imaging arms employ spatial light modulators for the all-digital control of the projections, being able to dynamically change the measuring technique and the spatial properties of the states under study. Finally, we provide a detailed theory that explains the reported observations.
\end{abstract}

\vspace{2pc}

\submitto{\NJP}

\section{Introduction}

Ghost imaging was first performed by Pittman et.~al \cite{pittman1995optical}, in which entanglement was utilized as the source of spatial correlations between a pair of separate photons. In quantum ghost imaging, one photon of the pair interacts with an arbitrary object and is collected with a bucket detector with no spatial resolution. The other photon, in the imaging arm, does not interact with the object but rather is sent directly to a spatially-resolving device for detection, usually a 2D scanning detection system or a camera. Despite neither photon being able the recover the shape of the object by itself, an image can be reconstructed when measuring in coincidences due to the spatial correlations created prior to the interaction with the object, i.e., within the nonlinear crystal.

The first ghost imaging tests made use of entanglement as the source of spatial correlations, such as those arising from the spontaneous parametric down-conversion (SPDC) process \cite{walborn2010spatial}. However, classical intensity correlations from a thermal light source have also been used to demonstrate ghost imaging \cite{bennink2002two, bennink2004quantum, valencia2005two}, showing the analogy between the two scenarios \cite{gatti2004ghost, cai2005ghost}. Subsequently, ghost imaging has been studied from a computational perspective (a technique which only requires bucket detectors) \cite{erkmen2010ghost, shapiro2008computational} and using compressive sensing to reduce the number of required measurements \cite{katz2009compressive}. Ghost imaging has also been observed in various degrees of freedom (DoF), such as the orbital angular momentum of light  \cite{jack2009holographic}, correlations in the time domain \cite{ryczkowski2016ghost}, in momentum-position \cite{howell2004realization} and spectral DoF \cite{chan2009two}. 3D ghost images have been reconstructed using single-pixel detectors~\cite{sun20133d}, and ghost imaging has even been studied in the presence of turbulence \cite{cheng2009ghost}. See Refs. \cite{shapiro2012physics, moreau2019imaging} for comprehensive reviews. Recently, the concept of ghost imaging was extended to entanglement swapped photons, demonstrating ghost imaging with initially independent photons \cite{bornman2018contrast}.  In this case the role of state symmetry was crucial to the outcome of the object/image contrast.

Here we demonstrate a new form of ghost imaging where the object and image arms are placed after a Hong-Ou-Mandel (HOM) interference filter \cite{hong1987measurement}, allowing the biphoton imaging to be carried out using either symmetric or anti-symmetric states. Furthermore, we employ spatial light modulators (SLMs) to dynamically control both the object and image, in particular, using digitally-controlled holograms on the image arm to reconstruct the object without a mechanical scanning system or a spatially-resolved camera.  We show that our ghost imaging setup including the HOM filter results in the reconstruction of an image comprised of a `double object', with each reconstructed object rotated in opposite directions. This is explained by the action of the symmetry selection step comprising a beamsplitter (BS) and Dove prisms.

\begin{figure}[t]
\centering
\includegraphics[width=\linewidth]{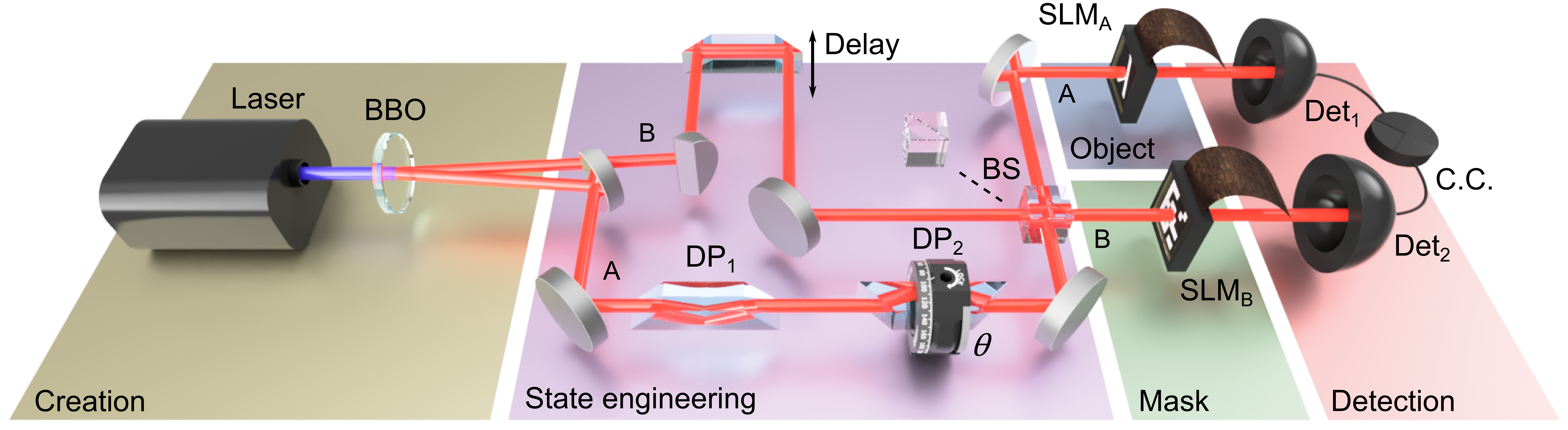}
\caption{Detailed experimental setup description, comprising the creation (golden), state engineering (purple), object (blue), mask (green) and detection (red) steps. BBO: nonlinear crystal; DP$_{1\&2}$: dove prisms; BS: 50:50 beamsplitter; SLM$_{\text{A\&B}}$: spatial light modulators; Det$_{1\&2}$: bucket detectors formed by interference filters, few-mode fibres and avalanche photo-diodes; C.C.: coincidence counter.}
\label{fig:setup}
\end{figure}

\section{Experiment}

We start describing the experimental setup in Fig.~\ref{fig:setup} to easily identify the role of each optical element involved later on in the Theory section. The experiment is divided conceptually into three sections. In the first, an entangled biphoton state is produced using a SPDC photon pair source, resulting in a state that is always symmetric. In the second, we pass the photon pair through a quantum state engineering system comprising Dove prisms (to control state phases $\theta$) and a HOM interference filter to single out specific states based on their symmetry. Finally, in the third part we perform ghost imaging using the engineered two photon state, consisting of the object and mask projections and photon pair detection. A detailed description of the experimental setup is given in the Supplementary Material.

We employ tools common in computational ghost imaging, namely digital projections for the image reconstruction, allowing the use of two bucket detectors and removing the need for cameras or mechanical scanning systems. 
To perform the ghost imaging measurements, the binary object, $O$, that we wish to reconstruct is encoded on SLM$_\text{A}$, and the scan is performed by dynamically modifying the hologram encoded on SLM$_\text{B}$. The different procedures used to reconstruct the image, single pixel and random mask scans, are introduced in the Results section.

\section{Theory}
\label{sec:theorysection}

Spatially-entangled photon pairs are generated in the nonlinear crystal (BBO). After propagating along the optical elements comprising the symmetry filter, the photons of each pair, A and B, are sent to the SLM screens. SLM$_\text{A}$ is masked with a binary object $O$ of our choosing, and SLM$_\text{B}$ is used to perform measurements. Based on said measurements on photon B, $O$ can be reconstructed when detected in coincidence with photon A.

To study the effect of state symmetry on the reconstructed object, we first study the setup using the orbital angular momentum (OAM) basis of the photons~\cite{allen1999iv}. Any set of spatial modes which form a basis can be used to express a mode of light with an arbitrary spatial profile, e.g. the Laguerre-Gaussian (LG), or Hermite-Gaussian (HG) modes. It is also evident that any arbitrary state can be written as the sum of a symmetric part and an anti-symmetric part. The effect that a state symmetry has on, for example, coincidence events in an entanglement experiment has recently been studied \cite{zhang2016engineering}, where it was shown how to control the spatial state symmetry by exploiting an HOM interferometric measurement \cite{hong1987measurement}, also known as an HOM filter. Such techniques work regardless of the spatial basis \cite{zhang2016hong}. The HOM filter passes only anti-symmetric states when conditioned on coincidences and the symmetry of the input state is tuned by adjusting the relative phases using two Dove prisms rotated by an angle of $\theta$ relative to one another. 

To begin, consider the state generated by SPDC at the crystal plane in the OAM basis

\be
\ket{\Psi} = \sum_{\ell}a_\ell\ket{\Psi^+_\ell},
\label{eqn:initialstateoam}
\ee
with $\ket{\Psi^+_\ell} = \frac{1}{\sqrt{2}}\left\{ \ket{\ell}_A\ket{-\ell}_B + \ket{-\ell}_A\ket{\ell}_B \right\}$, and $a_\ell$ the appropriate amplitude. The presence of the Dove prisms at a relative angle $\theta$ in path A has the effect $\ket{\ell}_A \to \ket{\ell}_A e^{i2\ell\theta}$, in which case Eq.~(\ref{eqn:initialstateoam}) transforms to

\bea
\ket{\Psi} & \to & \sum_{\ell}\frac{a_\ell}{\sqrt{2}}\left( \ket{\ell}_A\ket{-\ell}_Be^{i2\ell\theta} + \ket{-\ell}_A\ket{\ell}_Be^{-i2\ell\theta} \right) \nonumber \\
& = & \sum_{\ell}a_\ell\left( \ket{\Psi^+_\ell}\cos(2\ell\theta) + i\ket{\Psi^-_\ell}\sin(2\ell\theta) \right).
\label{eq:angle}
\eea

When the relative angle is set to $\theta = \frac{\pi}{4}$, the only $\ket{\Psi^+_\ell}$ ($\ket{\Psi^-_\ell}$) terms that survive are those with $\ell$ even (odd). With this state passed through the HOM filter, only the anti-symmetric modes (i.e., $\ket{\Psi^-_\ell}$, those with odd $\ell$ values) remain when conditioned on coincidences after the filter \cite{zhang2016engineering}. All symmetric states are removed, since they result in no coincidences.  

One might ask whether such symmetry filtering holds when any DoF other than OAM is considered. Symmetry is an intrinsic property of a quantum state: a state which is (anti-)symmetric in one basis is (anti-)symmetric in all bases (see Supplementary Material). Hence, we can express a state in any basis we choose without affecting the symmetry. When considering quantum imaging of arbitrary images, in which information is encoded in the transverse position of every pixel the image is comprised, these pixels are most easily described using a transverse position vector. Hence, it is intuitive to describe imaging in the \textit{position basis}.  With this in mind, Eq.~(\ref{eqn:initialstateoam}) can be re-expressed as

\be
\ket{\Psi} = \sum_{\bm{r} \in \mathcal{S}}c(\bm{r})\ket{\bm{r}}_A\ket{\bm{r}}_B,
\label{eqn:stateinitialOAM}
\ee
where the sum runs over all SLM pixels, a set we call $\mathcal{S}$. We consider this discrete case since the SLM itself consists of discrete pixels. Here $c(\bm{r})$ is the probability amplitude for photons A and B to be found in the crystal plane at the transverse position $\bm{r} = (x,y)$; they have the same position since they originate at the same point in the crystal.

Photon A passes through two Dove prisms (which are initially set to have a relative angle of $\theta=0$).  Later, when one of the Dove prisms in path A is rotated at an angle $\theta$ with respect to the other, $R(2\theta)$ will represent a rotation of the transverse position of photon A (for a setup without the Dove prisms, or with $\theta = 0$, we have $R(2\theta) = \mathbb{I}$). The explicit $\theta$ dependence of $R$ is suppressed for brevity. Note also that we assume paths A and B have the same path length unless stated otherwise. Therefore at the BS plane Eq.~(\ref{eqn:stateinitialOAM}) becomes

\be
\ket{\Psi} \to \sum_{\bm{r}}c(\bm{r})\ket{R\bm{r}}_A\ket{\bm{r}}_B.
\label{eqn:bsplane}
\ee

In the absence of a BS and hence an HOM filter, the SLM is placed at the crystal plane and so our `no beamsplitter' state, $\ket{\Psi_{nbs}}$, at the SLM plane is

\be
\ket{\Psi_{nbs}} = \sum_{\bm{r}}c(\bm{r})\ket{R\bm{r}}_A\ket{\bm{r}}_B,
\label{eqn:nobs}
\ee
which shows a rotation of the transverse position of photons in path A. In such a case, it is predicted that the outcome will match that of a conventional ghost imaging experiment, save for the measured image being rotated by an angle of $2\theta$ relative to the object. This is a corollary of the main study.

\subsection{Ghost imaging with an HOM filter}
\label{subsec:bs+matching}

In the presence of a 50:50 BS for HOM interference, and accounting for the number of mirror reflections in each path, the action of the filter is

\bea
\hspace{-0.4cm}\ket{\bm{r}}_{A} \to \frac{1}{\sqrt{2}}\left[ \ket{\bm{r}}_A + \ket{\bm{r}}_B \right] \; ; \; \ket{\bm{r}}_{B} \to \frac{1}{\sqrt{2}}\left[ \ket{\bm{r}}_B - \ket{\bm{r}}_A \right],
\label{eqn:bstransformations}
\eea
so that our `beamsplitter' state, $\ket{\Psi_{bs}}$, is

\bea
\hspace{-1cm}\ket{\Psi_{bs}} & = & \frac{1}{2}\sum_{\bm{r}}c(\bm{r})\left[\ket{R\bm{r}}_A + \ket{R\bm{r}}_B \right]\left[\ket{\bm{r}}_B - \ket{\bm{r}}_A \right] \nonumber \\
\hspace{-1cm}& = & \frac{1}{2}\sum_{\bm{r}}c(\bm{r})\left[\ket{R\bm{r}}_A\ket{\bm{r}}_B - \ket{\bm{r}}_A\ket{R\bm{r}}_B + \ket{R\bm{r},\bm{r}}_B - \ket{R\bm{r},\bm{r}}_A \right].
\label{eqn:isbs1}
\eea

We post-select on coincidences, allowing us to drop the latter two terms in Eq.~(\ref{eqn:isbs1}), so

\be
\ket{\Psi_{bs}} = \mathcal{K} \sum_{\bm{r}}c(\bm{r})\left[\ket{R\bm{r}}_A\ket{\bm{r}}_B - \ket{\bm{r}}_A\ket{R\bm{r}}_B \right],
\label{eqn:statepostselect}
\ee
with $\mathcal{K}$ the normalisation constant.

A comparison of all the imaging scenarios will be easier if all $R$ dependence is moved to photon B. In the Supplementary Material we demonstrate how the rotational dependence can be shifted from photon A to photon B, substituting $R$ by $R^{-1}$, so Eq.~(\ref{eqn:statepostselect}) can be written as

\be
\ket{\Psi_{bs}} = \mathcal{K} \sum_{\bm{r}} c(\bm{r}) \ket{\bm{r}}_A \left[\ket{R^{-1}\bm{r}}_B - \ket{R\bm{r}}_B \right].
\label{eqn:isbs}
\ee

We therefore predict that ghost imaging with an HOM filter setup will produce a result consisting of a juxtaposition of the original object $O$ rotated by an angle $2\theta$, and $O$ rotated by $-2\theta$.

\subsection{Beamsplitter without an HOM filter}
\label{subsec:bs+mismatching}

To affect HOM filtering, it is experimentally necessary to make use of a BS and perfectly match the lengths of paths A and B. Photons A and B then have identical time stamps and are indistinguishable. All of this gives rise to the well-known `HOM dip'.

However, we wish to study the effect of turning off the HOM filtering, but leaving the BS in place. This is achieved by slightly increasing the length of path B by way of the translation stage (the delay in Fig.~\ref{fig:setup}) so that the difference in path length is larger than the coherence length of the SPDC detected photons. Photon B is ergo slightly delayed with respect to photon A and the photons are distinguishable. We indicate the presence of this time delay of photon B by means of a prime symbol, $\ket{\bm{r}}_B \to \ket{\bm{r}'}_B$. Effecting this change in photon B in Eq.~(\ref{eqn:bsplane}) while applying the BS transformations in Eq.~(\ref{eqn:bstransformations}), and thereafter post-selecting on coincidences, gives

\be
\ket{\Psi_{bs}'} = \mathcal{K} \sum_{\bm{r}}c(\bm{r}) \left[ \ket{R\bm{r}}_A \ket{\bm{r}'}_B - \ket{\bm{r}'}_A \ket{R\bm{r}}_B \right].
\label{eqn:ketbsnohom}
\ee

Be that as it may, since the object masking SLM$_\text{A}$ is static and the time taken for each step of the measurement protocol carried out using SLM$_\text{B}$ is orders of magnitude larger than the time taken for photon B to travel the extra distance of the mismatched path B, experimentally, the time delay of photon B cannot be observed. Therefore, results obtained for the mismatched path length case (i.e. with a non-zero $\theta$ and BS present, but no HOM filtering) appear identical to the HOM filtering case, so $\ket{\Psi_{bs}'} \equiv \ket{\Psi_{bs}}$.

\subsection{Object reconstruction}
\label{subsec:Object}

Given either engineered state $\ket{\Psi_{nbs}}$ or $\ket{\Psi_{bs}}$, the detection section of the experiment is carried out by masking SLM$_\text{A}$ with a binary object $O$, the information of which is contained in the function $O(\bm{r})$: $O(\bm{r}) = 0$ if the pixel at position $\bm{r}$ in SLM$_\text{A}$ is black in the object, and $1$ if pixel $\bm{r}$ is white. Here, black means the SPDC photons are blocked (or deviated from the optical axis to be more precise) and white means the reflected photons are properly detected. The operator describing this masking process is $\ket{O}_A = \mathcal{N}\sum_{\bm{r}}O(\bm{r})\ket{\bm{r}}_A$, with $\mathcal{N}$ the appropriate normalization. After masking SLM$_\text{A}$ with $O$ and absorbing $\mathcal{K}$ into $\mathcal{N}$, the state of photon B, in the absence of the beamsplitter, is

\be
\bra{O}\Psi_{nbs}\rangle = \mathcal{N}^*\sum_{\bm{r}}c(\bm{r})O(R\bm{r})\ket{\bm{r}}_B.
\label{eqn:statecnobs}
\ee

In the case of HOM filtering, as well as the case of a non-zero $\theta$ - BS combination but mismatched path lengths, the state is

\be
\bra{O}\Psi_{bs}\rangle = \mathcal{N}^* \sum_{\bm{r}} c(\bm{r}) \left[ O(R\bm{r}) - O(R^{-1}\bm{r}) \right]\ket{\bm{r}}_B.
\label{eqn:statecbs}
\ee

If we set the weighting coefficients $c$ to unity, we can visualize the outcome more clearly 

\be
\bra{O}\Psi_{nbs}\rangle \propto \sum_{\bm{r}}O(R\bm{r})\ket{\bm{r}}_B,
\label{eqn:statenobs}
\ee

\be
\bra{O}\Psi_{bs}\rangle \propto \sum_{\bm{r}} \left[ O(R\bm{r}) - O(R^{-1}\bm{r}) \right]\ket{\bm{r}}_B,
\label{eqn:statebs}
\ee

\noindent
where the operator $R = R(2\theta)$ is the rotation in the transverse plane. Both of these formulae match the earlier predictions, namely: a single image rotated relative to the object in the case of Eq.~(\ref{eqn:statenobs}), and a juxtaposed `double' image with opposite rotations in the case of Eq.~(\ref{eqn:statebs}). The intensity of pixel $\ket{\bm{r}}_B$ in the reconstructed object in each case is respectively

\be
\left| \bra{\bm{r}}_B\bra{O}\Psi_{nbs}\rangle \right|^2 \propto \left| O(R\bm{r}) \right|^2,
\label{eqn:simplerot}
\ee

\be
\left| \bra{\bm{r}}_B\bra{O}\Psi_{bs}\rangle \right|^2 \propto \left| O(R\bm{r}) - O(R^{-1}\bm{r}) \right|^2.
\label{eqn:doublerot}
\ee

Equations (\ref{eqn:simplerot}) and (\ref{eqn:doublerot}) are key to understanding the object reconstructions shown in the following section.

\section{Results and discussion}
\label{sec:resultsanddiscussion}

\begin{figure}[t]
\centering
\includegraphics[width=\textwidth]{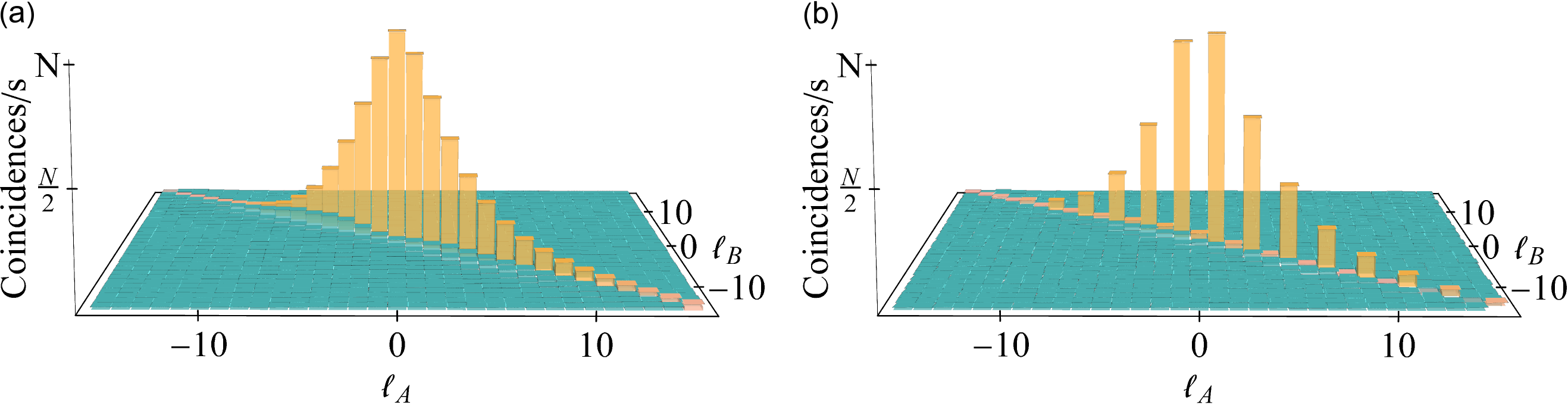}
\caption{Experimental symmetry spatial filter by means of an HOM measurement within the OAM topological charge range $\ell_A = \ell_B = [-15,15]$. (a) OAM spiral bandwidth of the SPDC photons when no HOM filtering is implemented, and (b) the analogous spiral bandwidth after introducing the HOM filter and setting the relative angle between the Dove prisms to $\theta = \frac{\pi}{4}$. The coincidence counts are normalized with respect to their respective maxima.}
\label{fig:dipspeaks}
\end{figure}

First we confirm the SPDC spiral bandwidth and the HOM filtering (the first two sections of the experiment in Fig.~\ref{fig:setup}), with the results given in Fig.~\ref{fig:dipspeaks}. Here, the OAM spiral bandwidth of the SPDC photons is experimentally measured within the range $\ell_A = \ell_B = [-15,15]$, with the data in Fig.~\ref{fig:dipspeaks}(a) taken without a BS, and that of Fig.~\ref{fig:dipspeaks}(b) taken after introducing a BS and setting $\theta = \frac{\pi}{4}$, forming an HOM filter.

In what is to follow, we analyze the most important experimental results as predicted in the theory section. We first give the reconstructed object obtained in a standard ghost imaging setup, but instead use the SLM to dynamically encode the masks needed for each measurement. Next we show the effect of rotating one of the Dove prisms with respect to the other, and finally we implement the HOM filter before performing ghost imaging. 

\subsection{Rotated ghost imaging reconstruction}
\label{subsec:ghostrotatedDP}

\begin{figure}[t]
\centering
\includegraphics[width=0.83\linewidth]{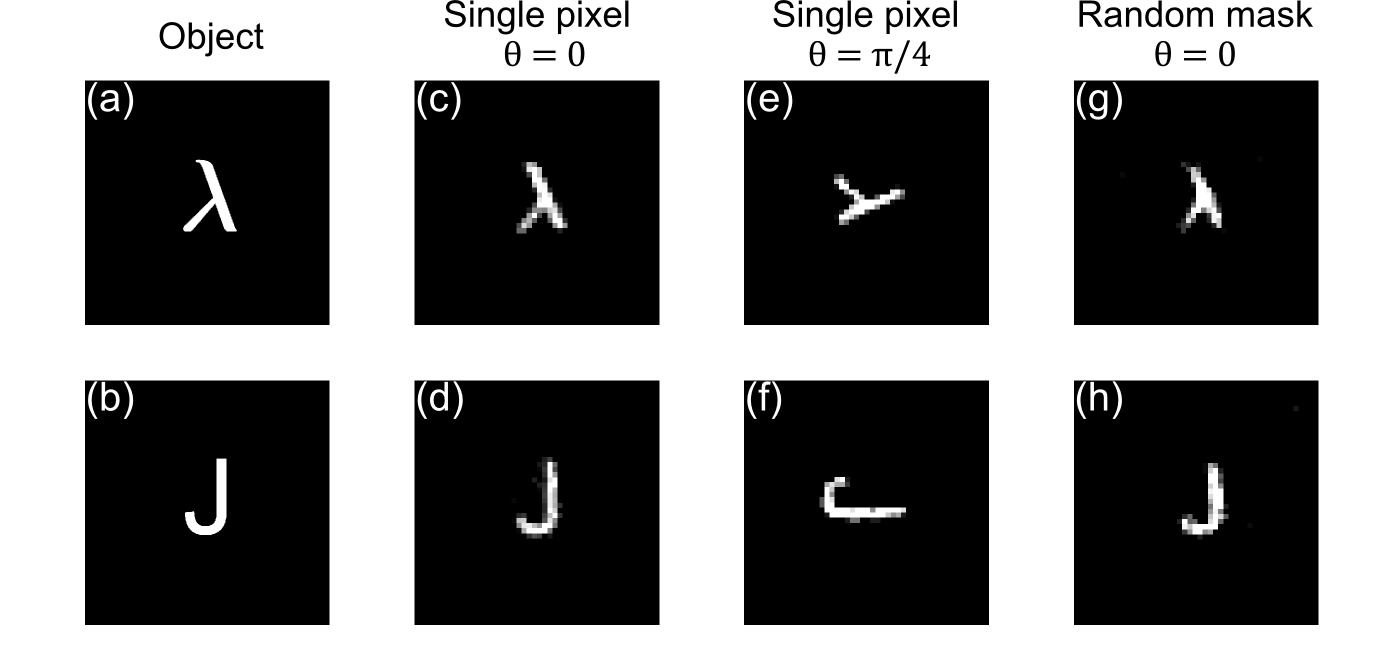}
\caption{Ghost imaging results, with no BS ($\ket{\Psi_{nbs}}$), using the SLM to encode the masks. (a,b) The objects $O$ encoded in path A, with white pixels indicating transmitted photons and the black pixels blocked photons. The reconstructed image of the corresponding object on the left, (c-f) using a single pixel 48$\times$48 scan and the relative Dove prism angle in Eq.~(\ref{eq:angle}) set to (c,d) $\theta = 0$ and (e,f) $\theta=\frac{\pi}{4}$, or (g,h) using a random mask scan with the same resolution and angle $\theta = 0$ as in (c,d).}
\label{fig:ghostslm}
\end{figure}

First, an experiment was run with the setup as depicted in Fig.~\ref{fig:setup}, but without the HOM filter (the BS was removed). The SLM in path A was masked with a 960$\times$960 resolution object $O$, as shown in Figs.~\ref{fig:ghostslm}(a,b), while performing a digital raster scan using the SLM in path B (with a  48$\times$48 resolution `on pixel'). The results are shown in Figs.~\ref{fig:ghostslm}(c,d) with a Dove prism angle of $\theta = 0$ and in Figs.~\ref{fig:ghostslm}(e,f) when $\theta = \frac{\pi}{4}$. The ghost images were reconstructed using the set of coincidence counts $\{ c_i \}$ for every raster position in SLM$_\text{B}$ as
\be
\text{Image} = \frac{c_1}{n}P_1 + \frac{c_2}{n}P_2 + \cdots,
\label{eqn:singlepixelimage}
\ee
where $c_i$ is the coincidence count recorded for raster position $P_i$, and $n$ is a normalization constant (see Supplementary Material). The results confirm the accuracy of the digital scan approach. However, the resolution that can be used in such a single `on pixel' reconstruction technique is limited by the strength of the signal arriving at the SLM. The integration time for each raster position increases as the pixel size decreases, in order to overcome the noise.

A different measurement scheme, a random mask scan \cite{chan2008single} based on the compressed sensing concept \cite{donoho2006compressed}, was also tested for the object reconstruction in order to overcome the noise in low signal cases without the need to decrease the resolution~\cite{sun2016improving}, as shown in the examples of Figs.~\ref{fig:ghostslm}(g,h). As before, SLM$_\text{A}$ is masked with a static 960$\times$960 binary object $O$. However, instead of scanning over every pixel in SLM$_\text{B}$ individually and recording the corresponding coincidence count, the random mask scheme involves first generating a set of $N$ random binary masks, with 50\% of the pixels white and 50\% of the pixels black, randomly so, for each mask. Then, SLM$_\text{B}$ is encoded with one of these random binary masks and the corresponding coincidence counts recorded. This process is repeated for every random mask. Finally, with the set of random binary masks $\{ M_i \}$ and their corresponding coincidence counts $\{ c_i \}$, for a large enough $N$, the object is reconstructed by again taking a convex combination of images, with the images in this scheme being the weighted random masks themselves, i.e.
\be
\text{Image} \approx \frac{(c_1 - \boldsymbol{\Bar{c}})}{n}M_1 + \frac{(c_2 - \boldsymbol{\Bar{c}})}{n}M_2 + \cdots,
\label{eqn:randommaskimage}
\ee
where $c_i$ is the coincidence count recorded for each random mask $M_i$, $n$ is a normalization constant, and $\boldsymbol{\Bar{c}}$ is the average of all coincidence counts measured~\cite{sun2016improving}. This is done since the `on' pixel would ordinarily correspond to a value of 1 and the `off' pixel to -1, giving an average outcome of 0. But in our case, the `off' pixel corresponds to 0, thus the non-zero average values must be subtracted to remove the noise. An animated example of the random mask reconstruction of Fig.~\ref{fig:ghostslm}(g) can be observed in the attached animation file (Lambda reconstruction), where the object is given in the leftmost, the real random mask used for each scan (iteration) is given in the middle, and the reconstructed image appears in the rightmost. The reconstructed image becomes clearer as the number of iterations, shown at the top, increase.

It is worth mentioning that this scheme can be generalised to cases with arbitrary proportions of black:white pixels. As we decrease the proportion of white pixels, we decrease the average of the measured coincidences which needs to be subtracted, i.e., the measurements are less noisy when not performing the average subtraction, with the extreme case being only 1 pixel as in Eq.~(\ref{eqn:singlepixelimage}). However, the maximum attainable resolution decreases, for a given signal arriving at the SLM, when decreasing the proportion of white pixels.

\begin{figure}[t]
\centering
\includegraphics[width=\linewidth]{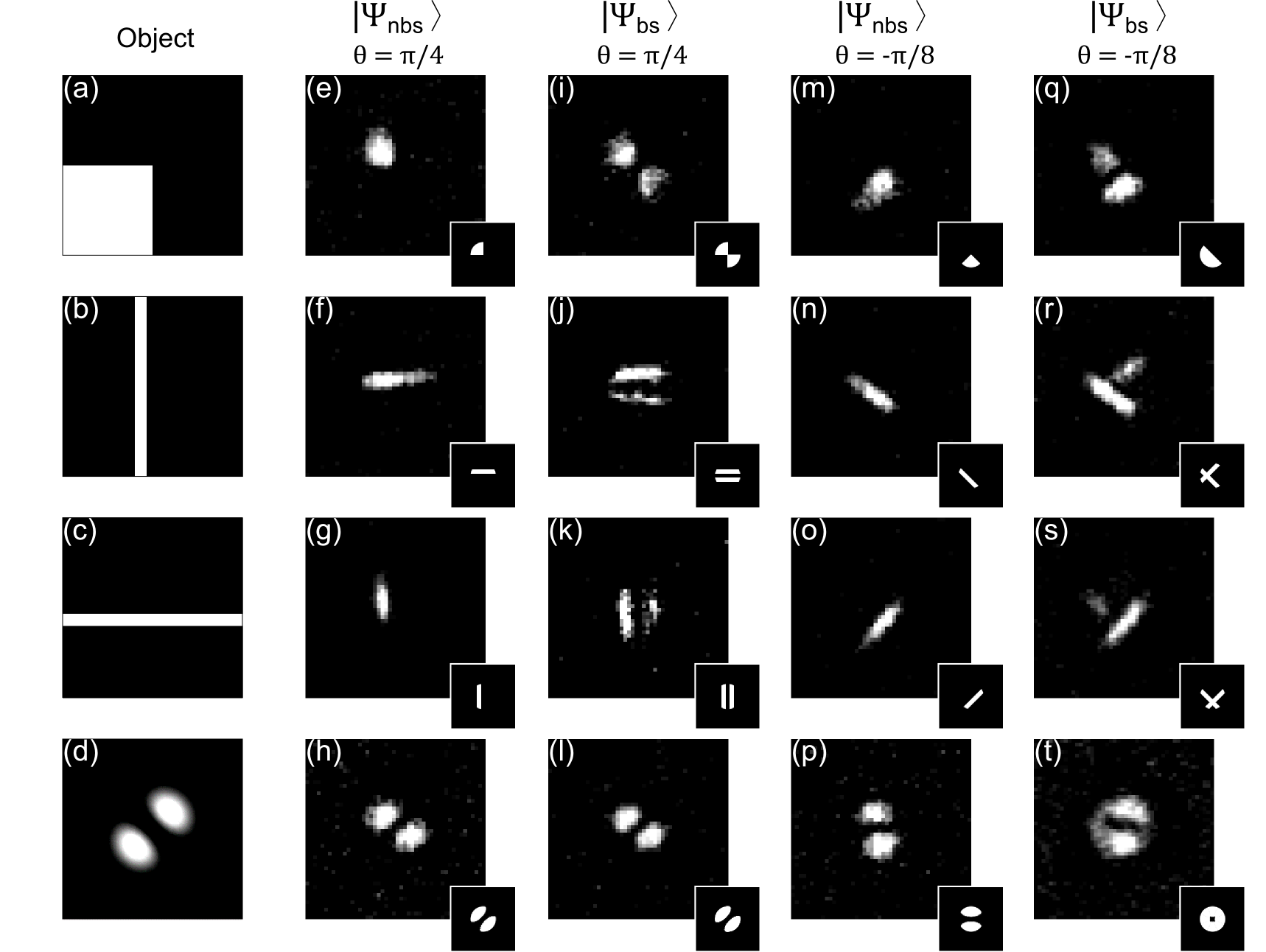}
\caption{Ghost imaging with and without HOM filtering, using the random mask scan sequence with 48$\times$48 resolution. (a-d) The objects $O$ encoded in SLM$_\text{A}$. (e-h, m-p) The reconstructed image results considering the objects on the leftmost without a BS and (e-h) $\theta=\frac{\pi}{4}$ or (m-p) $\theta=-\frac{\pi}{8}$ from Eq.~(\ref{eq:angle}). (i-l, q-t) The reconstructed image of the leftmost objects including a coherent superposition at the BS and (i-l) $\theta=\frac{\pi}{4}$ or (q-t) $\theta=-\frac{\pi}{8}$. Insets show the simulated results taking into account a perfect aligned projections and the overlap with the 5 mm diameter SPDC beam.} 
\label{fig:ghostanti}
\end{figure}

To test this measurement technique in a ghost imaging setup, the experiment was run with the objects given in Figs.~\ref{fig:ghostanti}(a-d), using $N=4000$ different random masks, recording the coincidences with an integration time of 1 second per mask, and setting the relative Dove prism angle to $\theta=\frac{\pi}{4}$ for the results in Figs.~\ref{fig:ghostanti}(e-h), and $\theta=-\frac{\pi}{8}$ for those in Figs.~\ref{fig:ghostanti}(m-p).

From these results, the reconstructed image is rotated by an angle of $2\theta$ with respect to the original object, as predicted in Eq.~(\ref{eqn:simplerot}). This confirms the effect of Dove prisms on ghost imaging and lends credence to the idea of performing such calculations in the chosen position basis.

\subsection{Double ghost images}
\label{subsec:ghostrandommasknobs}

Next, to implement an HOM filter and investigate its effect on the reconstructed image, the relative Dove prism angle was set to a non-zero value and a beamsplitter (BS) inserted into the setup, which selects the state $\ket{\Psi_{bs}}$. As per Eq.~(\ref{eqn:doublerot}), the intensity of pixel $\bm{r}$ in the reconstructed image is a combination of the intensity of pixel $\bm{r}$ in $O$, rotated by both $R(2\theta)$, and by $R^{-1}(2\theta) = R(-2\theta)$. As stated, the reconstructed image will hence be a juxtaposition of $O$ rotated by $2\theta$ and $O$ rotated by $-2\theta$. This is confirmed experimentally in Figs.~\ref{fig:ghostanti}(i-l) for a relative Dove prism angle of $\theta=\frac{\pi}{4}$, and in Figs.~\ref{fig:ghostanti}(q-t) for $\theta=-\frac{\pi}{8}$.

The experimental results in each row are for the objects given in the first column. Note that the results in the last row of Fig.~\ref{fig:ghostanti} are identical, with or without the beamsplitter and $\theta=\frac{\pi}{4}$, and match the intensity profile of the object, save for the rotation. In other words, we do not see the `double' image in the reconstructed images. This is a result of the original object being invariant under a rotation by $\pi$. This image invariance under rotations could play a role in future applications where the study of the innate geometric symmetry of an object is important, or it may find application in the field of quantum communication, wherein one could ascertain the centre of an SPDC beam source and align a system accordingly by using the counter-rotated reconstructed object.

Note that the experimental results slightly differ from their simulations shown in the insets, due to the difference in reflection/transmission ratios of the beamsplitter. We expect this to be the reason of the anti-clockwise-rotated portion of the reconstructed image to be dimmer compared with the clockwise-rotated portion; each half of the SPDC state, one in arm A and the other in arm B, traverses different ports of the beamsplitter. On the other hand, we deliberately displaced the object from the SPDC beam center of coordinates adding extra space between the reconstructed images, to properly identify the double rotation effect.

\begin{figure}[t]
\centering
\includegraphics[width=\linewidth]{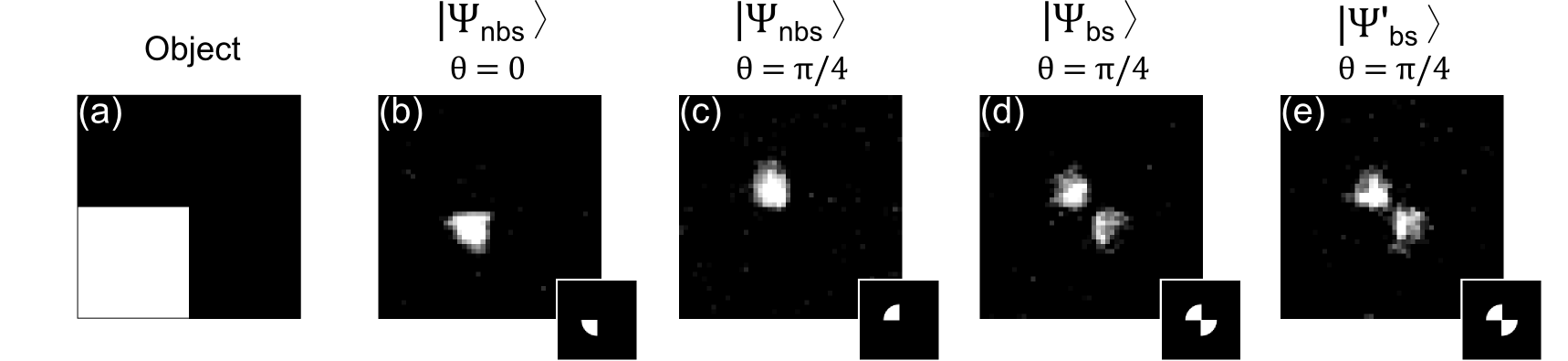}
\caption{Ghost imaging result summary. (a) Object; (b) no Dove prisms nor BS; (c) $\theta=\frac{\pi}{4}$ but no BS; (d) $\theta=\frac{\pi}{4}$ and HOM filter (BS present and path lengths matched), and (e) $\theta=\frac{\pi}{4}$, BS present, but with the length of path B slightly increased in order to obtain measurements without the symmetry filtering.}
\label{fig:ghostNoDip}
\end{figure}

Finally, Fig.~\ref{fig:ghostNoDip} gives a summary of all possible scenarios considered with the setup in Fig.~\ref{fig:setup}. In particular, the image in Fig.~\ref{fig:ghostNoDip}(e) was recorded after the length of path B was increased by 100 $\mu$m in order to remove the HOM effect but keeping the BS in. That is to say, Fig.~\ref{fig:ghostNoDip}(e) shows the results for the $\ket{\Psi_{bs}'}$ state. It was anticipated that $\ket{\Psi_{bs}} \equiv \ket{\Psi_{bs}'}$, which is confirmed experimentally given the fact that Figs.~\ref{fig:ghostNoDip}(d) and \ref{fig:ghostNoDip}(e) are qualitatively identical.

This image doubling can be understood as the beamsplitter `splitting' the image in two, and then being recombined after changing the path conditions. When measured in coincidence, a rotated photon A is either transmitted by the beamsplitter and interacts with the object, in which case the unrotated photon B (whose phase is $-2\theta$ with respect to photon A) is measured by the detection scheme, or the unrotated photon B is reflected by the beamsplitter and interacts with the object, with the rotated photon A (with a $2\theta$ phase relative to photon B) being measured.

Moreover, such `splitting' of the object into two rotated images is not restricted to any specific optical plane. This was tested by moving the BS to the Fourier plane of the crystal (and the SLM), with the results obtained in such a case identical to those reported here for the image plane.

\section{Conclusions}
We have used an HOM filter to engineer particular quantum states and used them in ghost imaging experiments. The results are in agreement with the theory and confirm the image rotation and image `doubling' as a consequence of the state preparation. Although such filtering is often understood in terms of the OAM basis, we translate it here to the position basis by virtue of the invariance of a quantum state's intrinsic symmetry under basis changes. In addition to an intriguing ghost imaging setup, we also employ all-digital control over the imaging arm for fast and convenient image reconstruction. Our work highlights important aspects of this form of ghost imaging and paves the way for further investigations and applications that employ imaging with specially engineered states.

\section*{Acknowledgements}

The authors would like to thank F.~S. Roux, M. Agnew, M.~J. Padgett and M.~Hendrych for valuable discussions. N.~B. acknowledges financial support from the South African CSIR DST-IBS programme and A.~V. from the Claude Leon Foundation.

\section*{References}

\bibliographystyle{iopart-num}
\bibliography{References_File}

\newpage
\clearpage

\section*{Supplementary Material}

\subsection*{Experimental setup}

The experimental setup is shown in Fig.~1 with the five parts comprising the creation (golden), state engineering (purple), object (blue), mask (green) and detection (red) steps. A mode-locked laser operating in the picosecond regime with 80 MHz pulse repetition rate, centred at a wavelength of 355 nm with an average power of 350 mW, is used to pump a 3-mm-long $\beta$-barium borate (BBO) nonlinear crystal. Spatially-entangled photon pairs centered both at a wavelength of 710 nm (labelled as A and B) are generated in the nonlinear crystal by means of a spontaneous parametric down-conversion (SPDC) type-I process. A small difference in the angle of emission existed between photons A and B (of around 3$^{\circ}$), making it easier to separate them with a D-shaped mirror. The photon in path A traverses two Dove prisms (DP$_1$ and DP$_2$). One of the Dove prisms is fixed to a rotation mount which is rotated by an angle $\theta$ about the optical axis to introduce a specific phase between spatial modes; the photon in path B is path length adjusted in order to achieve HOM interference.

The photons in A and B are then passed through a 50:50 beamsplitter (BS), the core element in the HOM filter. Only anti-symmetric input states result in a single photon in each arm, and so can be considered engineered when conditioned on coincidences. The symmetric states are tested by either removing the BS or working outside the HOM dip. Next, the photons are directed to the ghost imaging section of the setup: the object arm and image arm, the control of which are achieved using a single SLM encoded with amplitude and phase holograms (one half of the screen for the object and the other for the image). The plane of the crystal is relayed onto the SLM via paths A and B with a 5$\times$ magnification system (a 4$f$-system with f1 = 100 mm and f2 = 500 mm, not shown in Fig.~1), obtaining a $\sim$ 5 mm SPDC beam diameter at the SLM. The SLM plane is then relayed again with a 375$\times$ de-magnification system  (a 4$f$-system with f3 = 750 mm and f4 = 2 mm, also not shown) onto few-mode fibres (FMFs). FMFs (with $\sim13\,\mu$m core diameters) are used in order to increase the collection area as opposed to using single-mode fibres (with $\sim5\,\mu$m core diameters) and to reduce the noise which was observed when using multi-mode fibres (with $\sim62.5\,\mu$m core diameters). In combination with few-mode fibres, the SLM allowes for joint projective measurements of particular spatial modes to be made. Interference band-pass filters with bandwidths of 10 nm are used prior to the FMFs, which are in turn connected to avalanche photodiodes to detect the single photons, with coincidences registered (maximum never exceeded 50 KHz) via a coincidence counter. The coincidence window is set to 12 ns, avoiding any possible accidental coincidences (noise) from neighboring pulses.

\subsection*{$R/R^{-1}$ dependence shift between photons A and B}

We want to move the $R$ dependence from photon A to photon B in Eq.~(8) from the main text, so we can simplify the analysis by correlating the rotation effect only with photon B. With this in mind, we exploit the mathematical fact that, given a bijective mapping $\sigma: S_1 \to S_2$ from a finite set $S_1$ to a finite set $S_2$ with the same cardinality as $S_1$, the summation of a function of mapped elements of $S_1$, i.e. $\sum_{m \in S_1} f(\sigma(m))$, is equal to summation of the same function of elements in the mapped set $S_2$, $\sum_{n \in S_2}f(n)$. That is to say, since $\sigma(S_1) = S_2$, we have $\sum_{m \in S_1} f(\sigma(m)) = \sum_{m \in \sigma(S_1)}f(m)$. This idea can be extended to the following identity for arbitrary functions $f$ and $g$, where $\sigma^{-1}$ is the inverse bijection

\be
\sum_{m \in S_1} f(\sigma(m))g(m) = \sum_{m \in \sigma(S_1)}f(m)g(\sigma^{-1}(m)).
\label{eqn:permusum}
\ee

If we assume that the pixels of the SLM screen are small enough that every pixel in the transverse plane after the rotation $R$ can be associated, or `matched', with a unique pixel in the original, un-rotated plane, then $R$ is a bijection. In fact, $R$ is a special bijection, a \textit{permutation}, since the domain and range of $R$ are the \textit{same} set. Hence there exists an inverse permutation (rotation), $R^{-1}$, representing a rotation of the transverse plane by the same magnitude, but opposite direction, to $R$. Therefore, with $R^{-1}$ as the permutation and applying Eq.~(\ref{eqn:permusum}) to the first term in Eq.~(8) from the main text, we obtain

\be
\hspace{-0.2cm}\sum_{\bm{r} \in \mathcal{S}}c(\bm{r})\ket{R\bm{r}}_A\ket{\bm{r}}_B = \sum_{\bm{r} \in R^{-1}(\mathcal{S})} c(R^{-1}\bm{r})\ket{\bm{r}}_A \ket{R^{-1}\bm{r}}_B,
\ee
with $\mathcal{S}$ the set of discrete pixel positions. However, $R^{-1}$ is a permutation, or re-arranging, of the elements of $\mathcal{S}$. Since we are summing over all pixel positions, we can replace $R^{-1}(\mathcal{S})$ with $\mathcal{S}$ above, and considering that the generation coefficients are not affected by the rotation, we can write Eq.~(8) from the main text as

\be
\ket{\Psi_{bs}} = \mathcal{K} \sum_{\bm{r}} c(\bm{r}) \ket{\bm{r}}_A \left[ \ket{R^{-1}\bm{r}}_B - \ket{R\bm{r}}_B \right].
\ee

\subsection*{Image measurement}
\label{subsec:measurement}

Here we make further comments on the two measurement schemes employed in the main text.

\noindent\emph{Single pixel scan} -- Every pixel on the SLM screens has a unique position vector $\bm{r}$. Switching on the pixel in SLM$_\text{B}$ at position $\bm{r}_i$ is represented by the operator $\ket{M_i}_B = \ket{\bm{r}_i}_B$. The observable $|p_{nbs,i}|^2$ or $|p_{bs,i}|^2$, corresponding to Eq.~(11) or (12) from the main text, respectively (which can be taken to correspond with the number of coincidence counts per unit time), is then

\be
p_{nbs,i} = \bra{M_i}_B\bra{O}\Psi_{nbs}\rangle = \mathcal{N}^*c(\bm{r}_i)O(R\bm{r}_i),
\ee
\be
p_{bs,i} = \bra{M_i}_B\bra{O}\Psi_{bs}\rangle = \frac{\mathcal{N}^*}{2}\left[ c(\bm{r}_i) O(R\bm{r}_i) - c(R^{-1}\bm{r}_i)O(R^{-1}\bm{r}_i) \right].
\ee

Switching on every pixel in the measurement arm SLM in succession gives a set of observables, $\{|p_{nbs,i}|^2\}$ or $\{|p_{bs,i}|^2\}$, which are used to recreate the object $O$ according to $O(\bm{r}_i) = |p_{nbs/bs,i}|^2$.

\noindent\emph{Random mask scan} -- Create a set of $N$ random binary 50:50 black:white masks. For the $i$th mask in the set, define the operator representing the masking of the measurement arm SLM by $\ket{M_i}_B = w_i\sum_{\bm{s}}M_i(\bm{s})\ket{\bm{s}}_B$; the information about the $i$th mask is contained in the binary function $M_i$ (this scheme can be adapted to random masks with no restriction on the proportion of pixels which are `on' and `off', as long as all the masks conserves the same proportion; for a set of masks with a balanced proportion, as in our case, the coefficients $w_i$ will all be equal in magnitude, i.e, normalized with respect the same average measured coincidences, and can hence be absorbed into $\mathcal{N}$). Finally the observable $|m_{nbs,i}|^2$ or $|m_{bs,i}|^2$ is given by

\be
m_{nbs,i} = \bra{M_i}_B\bra{O}\Psi_{nbs}\rangle = \mathcal{N}^* \sum_{\bm{r}}c(\bm{r})O(R\bm{r})M_i^*(\bm{r}),
\ee
\be
m_{bs,i} = \bra{M_i}_B\bra{O}\Psi_{bs}\rangle = \frac{w_i^*\mathcal{N}^*}{2}\sum_{\bm{r}}c(\bm{r})\left[ O(R\bm{r})M_i^*(\bm{r}) - O(\bm{r})M_i^*(R\bm{r}) \right].
\ee

The object $O$, for a sufficiently large $N$, is reconstructed by way of the convex sum $O(\bm{r}) \approx \sum_{i=1}^N |m_{nbs/bs,i}|^2M_i(\bm{r})$.

\subsection*{The effect of a change of basis on a state symmetry}
\label{subsec:symmbasis}

A well-known result from high energy physics is that a change of basis does not change the nature of a state symmetric character. Here we outline a simple proof of this.

Firstly, let $\mathcal{H}_n := (V,\langle \cdot,\cdot \rangle)$ be a complex Hilbert space of dimension $n$, and let $\{ u_1, u_2, \cdots, u_n \}$, $\{ v_1, v_2, \cdots, v_n \}$ be two orthonormal bases of $V$. We define the linear operator `change of basis' matrix $M$ such that $Mu_i = v_i \; \forall \; i$. It is easy to see that $M$ is then unitary, i.e. $\langle Mx,My \rangle = \langle x,y \rangle \; \forall \; x,y \in V$. Let $x = \sum_{i=1}^n \alpha_i u_i$, $y = \sum_{j=1}^n \beta_j u_j$ be arbitrary, so

\be
\langle x,y \rangle = \sum_{i,j} \alpha_i \overline{\beta}_j \left\langle u_i, u_j \right\rangle = \sum_{i=1}^n \alpha_i \overline{\beta}_i,
\ee
\be
\langle Mx, My \rangle = \sum_{i,j} \alpha_i \overline{\beta}_j \left\langle Mu_i, Mu_j \right\rangle = \sum_{i,j} \alpha_i \overline{\beta}_j \left\langle v_i,v_j \right\rangle = \sum_{i=1}^n \alpha_i \overline{\beta}_i.
\ee

The converse (i.e. a unitary matrix is a change of basis matrix) can be shown too: let $U$ be a unitary matrix, let $\{ | u_i \rangle \}$ be an orthonormal basis, and let $| t_i \rangle := U| u_i \rangle$ for some set of vectors $\{ | t_i \rangle \}$. Then $\langle t_i | t_j \rangle = \langle u_i | U^{\dagger}U | u_j \rangle = \langle u_i | u_j \rangle = \delta_{i,j}$, so $\{ | t_i \rangle \}$ is an orthonormal basis too. Therefore, given a matrix $U$, $U$ is unitary iff $U$ represents a change of basis.

Next, let $\mathcal{H}_A, \mathcal{H}_B$ be Hilbert spaces of dimension $n$ and $m$ and let $\{ | i \rangle_A \}, \{ | j \rangle_B \}$ be respective orthonormal bases. Any arbitrary (anti-)symmetric state $\ket{\Psi} \in \mathcal{H}_A \otimes \mathcal{H}_B$ can be written as

\be
\ket{\Psi} = \sum_{i=1}^n \sum_{j=1}^m a_i b_j \left( \ket{i}_A\ket{j}_B + \nu \ket{j}_A \ket{i}_B \right),
\ee

\noindent
with $\nu = 1 (-1)$ for the (anti-)symmetric case. Next, define the \textit{exchange operator} $P$ which switches the two particles in a state $\ket{x_1,y_2}$: $P\ket{x_1,y_2} = \ket{y_2,x_1}$. If the state $\ket{x_1,y_2}$ is symmetric, then $\ket{x_1,y_2} = \ket{y_2,x_1}$; if it is anti-symmetric, then $\ket{x_1,y_2} = - \ket{y_2,x_1}$.

To apply $P$ to $\ket{\Psi}$ requires $\mathcal{H}_A = \mathcal{H}_B$, and hence $n = m$ (this is clearly in line with the requisite indistinguishably of the two particles; it doesn't make much sense to talk about symmetric/anti-symmetric states if the constituent particles are distinguishable), so

\bea
P\ket{\Psi} & = & \sum_{i,j=1}^n a_i b_j \left( \ket{j}_A\ket{i}_B + \nu \ket{i}_A \ket{j}_B \right) \nonumber \\
& = & \nu \sum_{i,j=1}^n a_i b_j \left( \ket{i}_A\ket{j}_B + \nu \ket{j}_A \ket{i}_B \right) = \nu \ket{\Psi}.
\eea

The eigenvalue of $P$, i.e., $\nu$, tells us whether the state $\ket{\Psi}$ is symmetric or anti-symmetric. Next, use two $n \times n$ unitary matrices $U_1$ and $U_2$ to change the bases of $\mathcal{H}_A$ and $\mathcal{H}_B$, respectively, to any other bases. It turns out that $P$ commutes with the change of basis transformation, $U_1 \otimes U_2$, if $U_1 = U_2$, so we have

\be
(U_1 \otimes U_1)P\ket{\Psi} = (U_1 \otimes U_1)(\nu \ket{\Psi}) = \nu (U_1 \otimes U_1)\ket{\Psi},
\label{eqn:1}
\ee

\noindent
and, since $U_1$ maps a basis to another basis

\bea
P(U_1 \otimes U_1)\ket{\Psi} & = & P\sum_{i,j=1}^n a_i b_j \left( U_1\ket{i}_A U_1\ket{j}_B + \nu U_1\ket{j}_A U_1\ket{i}_B \right) \nonumber \\
& = & \sum_{i,j=1}^n a_i b_j \left( U_1\ket{j}_A U_1\ket{i}_B + \nu U_1\ket{i}_A U_1\ket{j}_B \right) \nonumber \\
& = & \nu (U_1 \otimes U_1)\ket{\Psi}.
\label{eqn:2}
\eea

So, $P\ket{\Psi} = \nu \ket{\Psi} \implies P(U_1 \otimes U_1)\ket{\Psi} = \nu (U_1 \otimes U_1)\ket{\Psi}$, so the symmetry of a state $\ket{\Psi}$ is maintained by an arbitrary change of basis.

\end{document}